\documentclass[letter,conference]{IEEEtran}

\usepackage{amsmath,amsthm, amsfonts,amssymb,epsfig,subfigure}

\usepackage{graphicx}

\usepackage{color}

\newcommand{{\HC}}{{\textsc{Push \& Pull}}}

\newcommand{\ls}[1]
    {\dimen0=\fontdimen6\the\font
     \lineskip=#1\dimen0
     \advance\lineskip.5\fontdimen5\the\font
     \advance\lineskip-\dimen0
     \lineskiplimit=.9\lineskip
     \baselineskip=\lineskip
     \advance\baselineskip\dimen0
     \normallineskip\lineskip
     \normallineskiplimit\lineskiplimit
     \normalbaselineskip\baselineskip
     \ignorespaces}
\begin{document}

\title{P\&P protocol: local coordination of mobile sensors for self-deployment\\ \bigskip \emph{TECHNICAL REPORT}}
\author{
N. Bartolini, A. Massini, S. Silvestri\\
Department of Computer Science\\
University of Rome "Sapienza", Italy\\
{\normalsize \{bartolini, massini, simone.silvestri\}@di.uniroma1.it
}
}

\maketitle


\begin{abstract}
The use of mobile sensors is of great relevance for a number of
strategic applications devoted to monitoring critical areas where
sensors can not be deployed manually. In these networks, each sensor
adapts its position on the basis of a local evaluation of the
coverage efficiency, thus permitting an autonomous deployment.

Several algorithms have been proposed to deploy mobile sensors over
the area of interest. The applicability of these approaches largely
depends on a proper formalization of rigorous rules to coordinate
sensor movements, solve local conflicts and manage possible failures
of communications and devices.

In this paper we introduce P\&P, a communication protocol that
permits a correct and efficient coordination of sensor movements in
agreement with the PUSH\&PULL algorithm. We deeply investigate and
solve the problems that may occur when coordinating asynchronous
local decisions in the presence of an unreliable transmission medium
and possibly faulty devices such as in the typical working scenario
of mobile sensor networks.

Simulation results show the performance of our protocol under a
range of operative settings, including conflict situations,
irregularly shaped target areas, and node failures.
%
%
%
%
\end{abstract}


\section{Introduction}
The research in the field of mobile wireless sensor networks is motivated by
the need to monitor critical scenarios such as
wild fires, disaster areas, toxic regions or battlefields, where
static sensor deployment cannot be performed manually.

In these typical working situations, sensors may
be dropped from an aircraft or
sent from a safe location. In these cases the
initial deployment over the Area of Interest (AoI) is neither complete nor uniform
as would be necessary to  enhance the sensing capabilities and
extend the lifetime of the network.
Mobile sensors can dynamically adjust their position to improve coverage with
respect to their initial deployment. Sensor movements
should therefore be coordinated according to a distributed deployment algorithm.

Out of the solutions proposed in the literature so far for
mobile sensor deployment, those described in \cite{Zou2003,Heo2005,Chen03,Suckme2004}
are based on the virtual force approach which models the
interactions among sensors as a combination
of attractive and repulsive forces.
Other approaches are inspired by the physics of fluids and gases
such as \cite{Pac2006} and \cite{Kerr2004}.
Another methodology is based on the construction of Voronoi diagrams \cite{LaPorta06,LaPorta04}.
According to this proposal, each sensor iteratively calculates its own
Voronoi polygon, determines the existence of coverage holes
and moves to a better position if necessary.
The solutions proposed in \cite{DCOSS08} and \cite{mass_tech_report} provide instead
density driven actions to uniformly distribute sensors according to a regular grid pattern.

The applicability of these deployment algorithms largely depends on the proper
formalization of rigorous rules
to coordinate sensor movements,  solve local conflicts and manage
possible failures of communications and devices.

Previous proposals only focus on the design of distributed algorithms for the adaptive deployment
of mobile sensors, aiming at covering the area of interest according to given efficiency objectives,
in particular coverage completeness and uniformity and low energy consumption.
Seldom do previous works enter the details of the communication protocol necessary to
enable the application of the proposed algorithms.

The main contribution of this paper is a communication protocol
that defines the rules to deploy mobile sensors according to the  {\HC} algorithm proposed in \cite{mass_tech_report}.
This algorithm is based on the autonomic computing paradigm. It completely delegates to the single sensors every decision regarding movements and
action coordination.
This way self-organization emerges without the need of external coordination or human intervention as the sensors adapt their position on the basis of their
local view of the surrounding scenario.

Given the absence
of a centralized coordination unit, and the lack of synchronization, sensors
have a primary role in the realization of the algorithm actions. Therefore,
the design of the related coordination protocol is particularly challenging.

Indeed, under the execution of the \HC\ algorithm, several types of conflicts may occur
as several sensors often compete to cover the same position.
Sensors should be capable to solve such conflicts by means of only local interactions.
We deeply investigate and solve the problems that may occur when coordinating  asynchronous
local decisions in the presence of an unreliable transmission medium and possibly faulty devices that characterizes
the typical working scenario of mobile sensor networks.

The proposed protocol works in respect of the algorithm goals,
permitting  the realization of a complete and uniform stable coverage, with low energy consumption.
Simulation results show the performance of our protocol under a range of
operative settings,
including conflict situations, irregularly shaped target areas, and node
failures.


\section{Brief description of the Push \& Pull algorithm} \label{sec:the_idea}

\HC\ is a completely distributed algorithm \cite{mass_tech_report} for the realization of an autonomous  deployment of mobile sensors.
According to this algorithm sensors perform a complete coverage of the AoI by means of a hexagonal tiling.
Initially, several tiling portions are created concurrently and every sensor not yet involved in the creation of a tiling portion gives start to a portion of its own in an
instant randomly selected  over a given time interval.
In order to make the exposition clearer, we  outline
the algorithm, before giving details on the implementing protocol.

Let $V$ be a set of equal sensors endowed with location determination, boolean sensing and
isotropic communication capabilities.
Sensors are kept
in active mode for all the deployment phase.
The deployment consists in realizing a hexagonal grid with side length equal to
the {\em sensing radius} $R_\texttt{s}$.
This setting guarantees both coverage and connectivity
when $R_\texttt{tx} \geq \sqrt{3}R_\texttt{s}$.
A sensor which is deployed at the center of a hexagonal tile is called {\em snapped}.
$Hex(x)$ is  the hexagonal region whose center is covered by the snapped sensor $x$.
All the other sensors lying in $Hex(x)$ are called {\em slaves of $x$} and compose the set $S(x)$.
All sensors that are neither snapped nor slaves are called {\em free}.
The set composed by the
free sensors located in radio proximity to  $p$ and by its slaves is denoted by $L(p)$.
In the following, $s_\texttt{init}$ denotes any of the starter sensors.

The four main activities of the algorithm, {\em Snap, Push, Pull, Merge}, are executed in
an interleaved manner as described in the following paragraphs.

\noindent
{\bf Snap activity. }
At the beginning, each sensor may act as starter of a snap activity from
its initial location at an instant randomly chosen over a given time interval.
Sensor $s_\texttt{init}$  elects its position as the center of the first hexagon
of its tiling portion and changes its status to snapped.

Any just snapped sensor $p$ performs a
 {\em neighbor
discovery},
that allows it to gather
information regarding the sensors belonging to $L(p)$.
Among these, $p$ selects at most six sensors
to make them snap to the
center of adjacent hexagons.

A snapped sensor leads the snapping of as many slaves as possible. If all  the hexagons adjacent to $Hex(p)$ have
been covered, $p$ stops any further snapping.
If, after the completion of the snap action, the snapped sensor has still some slaves in its hexagon,
it gives start to
the push activity.
Otherwise, if some hexagons are left uncovered
because no more sensors in $L(p)$ are available,
$p$ starts the pull activity.

Such deployed sensors, in their turn, give start
to an analogous selection and snap activity,
thus expanding the boundary of the current portion.
This process goes on until no other snaps are
possible, because either the whole AoI is covered,
or  all sensors located at boundary tiles do not
have any un-snapped sensor to snap.

\noindent
{\bf Push activity. }
Snapped sensors, after the completion of their snapping activity,
may still be surrounded by un-snapped sensors located inside their hexagon.
In this case, they proactively push such
un-snapped sensors towards lower density areas
located
within their transmission range.

Given two snapped sensors $p$ and $q$ located in
radio proximity from each other,
$p$ may offer one of its slaves to $q$ and push
it inside the hexagon of $q$ if $|S(p)| \geq
|S(q)|+1$.

The push activity is allowed in the only directions
that verify the {\em Moving Condition} according to which the movements of sensors from
$Hex(p)$ to $Hex(q)$ are restricted to the only cases in which:

{\small $$\{|S(p)| > |S(q)|+1 \} \textrm{ }\vee \textrm{ }
\{ |S(p)| = |S(q)|+1  \textrm{ } \wedge \textrm{
} ord(p)>ord(q) \}.
$$}
where $ord(\cdot)$ is a function
initially set to the unique identity code of the
sensor radio device.

In order to avoid inconsistencies the snapped sensors involved in a push activity
always advertise their neighborhood of the
changes in the number of slaves as
if the ongoing movements were already concluded.

%

\noindent
{\bf Pull activity. }
Snapped sensors may detect a coverage hole adjacent to their hexagon
and not have available sensors to snap. In this
case, they send hole trigger messages,
so reactively attracting un-snapped sensors and making them fill the hole.

Namely, let $p$ be a snapped node detecting a hole in an adjacent hexagon, with $L(p)=\emptyset$.
If $p$ has not the possibility to receive any
slaves from its neighbor hexagons, i.e. the
Moving Condition is not verified for any of them,
then it activates the following trigger mechanism.

Sensor $p$ temporarily alters the value of its $ord$ function
to 0 and notifies its neighbors of this change by
means of a {\em trigger notification message}.
This could be sufficient to make the Moving
Condition true with at least a snapped neighbor,
so $p$ waits until either a new slave comes into
its hexagon or a timeout expires.
If a new slave enters in $Hex(p)$, $p$ sets back
its  $ord$ value and snaps the new sensor,
filling the hole.
If the timeout expires and the hole has not been
covered yet, the trigger mechanism is extended by
forwarding the trigger message to the
adjacent hexagons of $p$, whose snapped sensors
set their  $ord$ value to 1.
This mechanism is iterated by $p$ over snapped
sensors at larger and larger distance in the tiling
until an available slave is attracted and the hole is covered.

This way, each snapped sensor
involved in the trigger notification mechanism sets
its  $ord$ value
proportional to the distance from $p$. All the
timeouts related to each new forwarding are set
proportionally to the  distance reached by
the trigger mechanism.
At the expiration of the trigger timeout, each
involved node sets back its  $ord$ to the
original value.

Observe that, the detection of several holes may cause the same
sensor to receive more than one trigger
 message that it stores in a
pre-emptive priority queue, giving precedence to
the messages related to the closest hole.

\noindent
{\bf Tiling merge activity. }
The possibility that many sensors act as starters may temporarily lead to the creation
of
several tiling portions with different orientations.

Algorithm \HC\ provides a mechanism to merge all
these tiling portions into a unique regular
and uniformly oriented tiling.
When the boundaries of two
tiling portions come in radio proximity with each
other, the one which was started
first absorbs the other  by making its snapped
sensors move into more appropriate snapping
positions.

\medskip

The combination of the described activities expands the
tiling and, at the same time, does its best to
uniformly distribute redundant sensors over the
tiled area, avoiding oscillations.

\section{The sensor coordination protocol P\&P}
The implementation of the \HC\ algorithm requires the definition of a protocol for
the local coordination of the sensor activities.

The coordination protocol provides the rules to
solve
contentions that may happen in several cases. For
example, two or more snapped sensors can decide
to issue
a snap command to different sensors towards
the same hexagon tile or the same low density hexagon
can be selected
by several snapped sensors as candidate for receiving redundant slaves.
These contentions are solved by properly
scheduling actions according to message
time-stamps and by advertising
related decisions as soon as they are made.
The P\&P protocol is designed to minimize energy
consumption entailing a small number of message
exchanges,
which is possible because the algorithm decisions
are only based on a small amount of local
information.
Furthermore, we assume that P\&P works over a
communication protocol stack which handles possible transmission errors and message losses by means of timeout and retransmission mechanisms.
Therefore the treatment of occasional message losses at the underlying protocol level
implies the occurrence of delays in the corresponding messages at the P\&P level that are
dealt by P\&P with proper timeout mechanisms.

Before we enter the details of the protocol we introduce some definitions.
Remember that sensors may be in one of the following state: {\em snapped}, {\em free} or {\em slave}.

The \emph{real cardinality} of a snapped sensor $p$ is the number of slave sensors actually located inside $Hex(p)$, that $p$ can utilize to
perform the snap, push and pull actions. The \emph{virtual cardinality} of $p$ differs from the previous one as it is calculated considering all the ongoing snap, push, pull actions as if they were already concluded.
The set $\textrm{VP}(p)$ of vacant positions detected by sensor $p$ contains the centers of hexagons adjacent to $Hex(p)$ that are not yet occupied by any snapped sensor.

Table \ref{tab:messaggi} contains a summary of the message types used by protocol P\&P.

\begin{table*}[t]
\centering
\begin{tabular}{|l|l|}
\hline
{\bf Message name} & {\bf Message fields} \\
\hline \hline
\texttt{IAS} & ID, coordinates, starter timestamp  \\
\hline
\texttt{InfoSnapped} & ID, coordinates, virtual cardinality \\
\hline
\texttt{InfoSlave} & ID, coordinates, energy level \\
\hline
\texttt{InfoFree} & ID, coordinates \\ \hline
\texttt{SIP} & ID, receiver ID, target position coordinates\\ \hline
\texttt{AckSIP} & ID, receive ID \\ \hline
\texttt{ClaimPosition} & ID, coordinates, timestamp \\ \hline
\texttt{PositionTaken} & ID, coordinates \\ \hline
\texttt{InfoStopped} & ID, coordinates \\ \hline
\texttt{IAYS} & ID, receiver ID \\ \hline
\texttt{CardinalityInfo} & ID, virtual cardinality \\ \hline
\texttt{Offer} & ID, receiver ID, virtual cardinality, transaction ID \\ \hline
\texttt{AckOffer} & ID, receiver ID  \\ \hline
\texttt{MoveTo} &   ID, receiver ID, destination coordinates, destination snapped sensor ID, transaction ID\\ \hline
\texttt{InfoArrived} & ID, receiver ID, transaction ID, energy level \\ \hline
\texttt{HoleInfo} & ID, hop counter, order value, hole coordinates, timeout \\ \hline
\texttt{Subst} & ID, receiver ID, energy level  \\ \hline
\texttt{AckSubst} & ID, receiver ID  \\ \hline
\texttt{SubstArrival}  &  ID, receiver ID  \\ \hline
\texttt{ProfilePacket}  & ID, receiver ID, order value, priority queue, neighborhood information\\ \hline
\texttt{MoveToSubst} & ID, receiver ID, order value, priority queue, neighborhood information\\ \hline
\texttt{Retirement} & ID, hole coordinates\\
\hline
\end{tabular}
\caption{Summary of P\&P messages}
\label{tab:messaggi}
\end{table*}

\section{P\&P: snap activity}\label{sec:snap}

In order to
describe the snap activity, we need to distinguish three cases, according to the role
of the involved sensor.
Indeed the actions undertaken by the starter sensors, the already snapped sensors and the sensors
being snapped, are substantially different.

%
%
%

\subsection{Starter sensor behavior}

At the beginning, any sensor $p$
may give start to the creation of a tile portion by snapping itself to its present position
in an instant of time $t_\texttt{start}(p)$ randomly selected over a time interval of length $R_\texttt{tx}/v$, where $v$ is the sensor movement speed.
If at the instant $t_\texttt{start}(p)$, sensor $p$ has not yet received any message, it elects its position
as the center of the first hexagon and establish the orientation of its tile portion.
At this point $p$ executes the snap actions under the role of snapped sensor, as described in the following paragraph.

\subsection{Snapped sensor behavior} \label{sec:snapped_in_snap}

\subsubsection{Neighbor Discovery}
A snapped sensor $p$ broadcasts a \texttt{IAS} (I Am Snapped) message to perform a neighbor discovery.
Such message contains the ID of the sender snapped sensor, its geographic coordinates and the timestamp
of the starter action.
All sensors located in radio proximity to $p$ (with the exception of those slaves located in different
hexagons) reply to its \texttt{IAS}, with a message containing  role dependent information:
the snapped sensors reply with an \texttt{InfoSnapped} message, while the slave and the free sensors reply with
an \texttt{InfoSlave} and an \texttt{InfoFree} message respectively.
These replies contain key information to perform the \HC\ algorithm,
and in particular: all three types of replies contain the ID and geographic coordinates of the replying sensors,
while the \texttt{InfoSnapped} and \texttt{InfoSlave} messages contain additional information.
In particular, the \texttt{InfoSnapped} message includes also the virtual cardinality of the replying snapped sensors while
the \texttt{InfoSlave} message includes the energy level of the replying slave sensors.

Thanks to the execution of the neighbor discovery phase, a snapped sensor $p$ is informed regarding the presence of
vacant positions, i.e. knows the composition of VP$(p)$.

\subsubsection{Snap into position}
A snapped sensor $p$ selects the closest
sensor in $L(p)$ to
each uncovered position and sends it a \texttt{SIP} (Snap Into Position) message.
This message contains the target position of the correspondent snap action, and the ID
of the selected sensor.

If a sensor receives a \texttt{SIP}, and is available to fill the vacant position, it replies with an
\texttt{AckSIP} message. This message contains the ID of the sensor that received the \texttt{SIP}, necessary for $p$ to discriminate among the several sensors to which it sent \texttt{SIP} messages.
If a sensor receives a \texttt{SIP} when it is not available to fill the vacant position (e.g it has already been contacted by another sensor), it does not reply to the \texttt{SIP} message of $p$ and lets the \texttt{AckSIP} timeout expire. This way $p$ will be capable to select a new
sensor to snap in such still vacant position.

After the transmission of the \texttt{SIP} messages and the reception of the related \texttt{AckSIP}, $p$ updates its local information, i.e. the number of free sensors located within its transmission range and its \emph{virtual cardinality}.
This way it keeps into account the departure of some sensors from either its transmission range or its hexagon.

In order to update the information related to the snapped neighbors, $p$ waits for the reception of the corresponding
\texttt{IAS} messages, to be sure that position conflicts are solved (see \ref{sec:resolution}).
No messages are involved in this phase that
consists in a mere calculation based on locally available information.

Let $p$ be the sensor that is performing the snap action and let $q$ be the one to which $p$ sent a \texttt{SIP} message for
the position $x$.
Five cases may occur, described as follows.\\
\noindent 1) Sensor $p$ receives both the \texttt{AckSIP} and the \texttt{IAS} message from $q$. This means that the snap action performed by $p$ was successful, therefore $p$ can update the local information regarding the snapped neighbor $q$.

\noindent 2) Sensor $p$ receives the \texttt{AckSIP} from $q$ acknowledging its availability to fill position $x$, but a conflict occurs solved in favor of another sensor $r$, which reaches position $x$ before sensor $q$.
Hence $p$ receives an \texttt{AckSIP} from $q$ and a \texttt{IAS} from $r$ for the same position $x$. Thus
$p$ can update the local information regarding the snapped neighbor $r$.

\noindent 3) Sensor $p$ receives the \texttt{AckSIP} from $q$ acknowledging its availability to fill position $x$, but a failure occurred and the \texttt{IAS} timeout expires. If $p$ detects the availability of another sensor in $L(p)$ that can be snapped to position $x$, it retries the snap action. If such sensor is not available, $p$ starts the pull action.

\noindent 4) Sensor $p$ does not receive the \texttt{AckSIP} from $q$, but receives a \texttt{IAS} message for position $x$ from another sensor $r$, before the expiration of the \texttt{AckSIP} timeout. Sensor
$p$ can update the local information regarding the snapped neighbor $r$.

\noindent 5) Sensor $p$ does not receive the \texttt{AckSIP} from $q$ nor the \texttt{IAS} from any other sensor within the \texttt{AckSIP} timeout. If $p$ detects the availability of another sensor in $L(p)$ that can be snapped to position $x$, it retries the snap action. If such sensor is not available, $p$ starts the pull action.

\begin{figure}[t]
\begin{center}
\scalebox{0.5}{\includegraphics[]{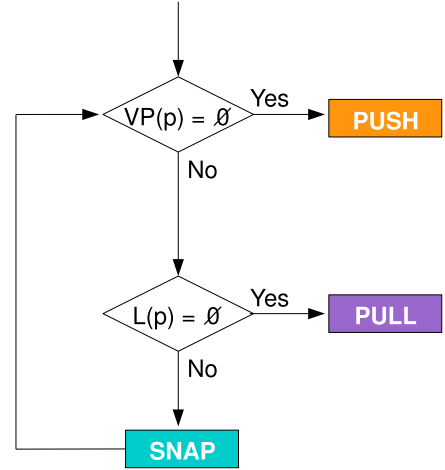}}
\end{center}
\caption{Snapped sensor behavior} \label{fig:snap_flow}
\vspace{-0.5cm}
\end{figure}

The behavior of a snapped sensor $p$ can be sketched as in Figure \ref{fig:snap_flow}.
It dwells in the snap phase until there are available sensors in $L(p)$ and vacant positions in the adjacent hexagons. If there are vacant positions and no sensors in $L(p)$, sensor $p$ gives start to the pull action
to attract new sensors from overcrowded areas.
If otherwise there are available sensors in $L(p)$ and no vacant positions, $p$ starts the push action
to uniform the redundant sensor distribution.

\begin{figure}[t]
\begin{center}
\scalebox{0.5}{\includegraphics[]{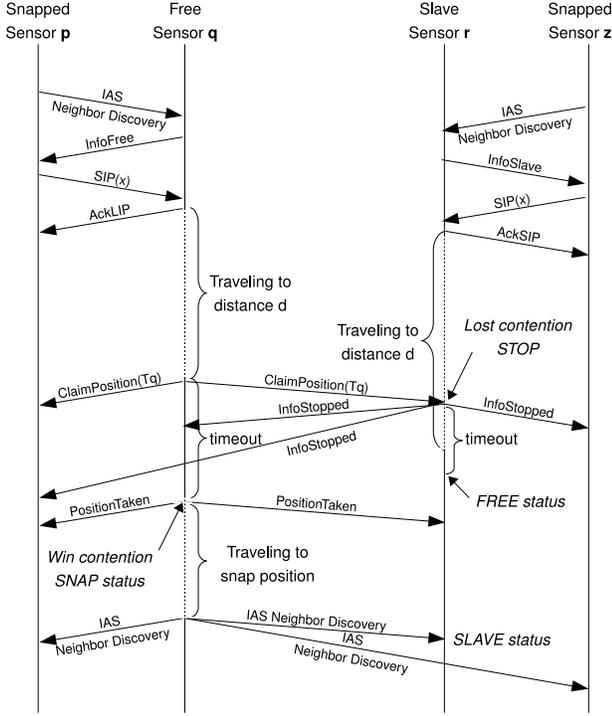}}
\end{center}
\caption{A typical scenario of snap position conflict between two
sensors} \label{fig:conflict} \vspace{-0.5cm}
\end{figure}

\subsection{Behavior of the sensors being snapped}

\subsubsection{Sensor localization}
A free sensor $q$ which receives a \texttt{IAS} message, coming from a snapped sensor $p$, replies with either an \texttt{InfoFree} or an \texttt{InfoSlave}
message depending on its position with respect to $p$.
If $q$ is located outside the hexagon of $p$, it remains in the free state and replies
to $p$ with an \texttt{InfoFree} message.
If instead $q$ is located inside the hexagon of $p$, it switches its state to slave
and replies to $p$ with an \texttt{InfoSlave} message.
In both cases $q$ becomes part of the set $L(p)$, that is the set of sensors that $p$
can snap to its adjacent vacant positions.
Notice that if $q$ is a slave, there is only one snapped sensor $p$ such that
$q \in L(p)$, thus slaves belonging to already snapped sensors do not reply to the \texttt{IAS} message of $p$.
If instead $q$ is a free sensor, it may belong to several sets $L(\cdot)$, for
different snapped sensors located in radio proximity from $q$ itself.

\subsubsection{Snap into position}
Sensor $q$, be it free or slave, at a certain time, may receive a \texttt{SIP} message coming from a snapped sensor.
Slaves reply only to \texttt{SIP} messages coming from their related snapped sensor, while free sensors only reply the first \texttt{SIP} message they receive and ignore subsequent ones.

After sending the \texttt{AckSIP} reply, sensor $q$ travels towards the snapping destination until it reaches a distance $d$ from it. Distance $d$ is set small enough to guarantee the
radio connectivity within the circular disk of radius $d$ and the inclusion of such disk into the hexagonal tile. Therefore $d \leq \sqrt{3} R_\texttt{s}/2$.

At this point sensor $q$ stops and broadcasts a  \texttt{ClaimPosition} message containing a timestamp
and waits for the expiration of a timeout to evaluate if other sensors are trying to
snap in the same position and in case to resolve the related contention.
At the timeout expiration, if no conflicts occurred or if a conflict was solved in its favor, $q$  switches its state to snapped,
sends a \texttt{PositionTaken} message and
 proceeds towards the destination.
After being  successfully snapped, sensor $q$ starts its own snap activity.

\subsubsection{Resolution of snap position contention} \label{sec:resolution}
Three events may occur when one or more sensors are engaged in a conflict with sensor $q$ due to the
contention for the same snap position: \\
1) sensor $q$ receives a \texttt{ClaimPosition} or a \texttt{PositionTaken} before reaching distance $d$ from the destination,\\
2) sensor $q$ receives a \texttt{ClaimPosition} after the arrival at distance $d$ from the destination and
before the expiration of the related timeout,\\
3) sensor $q$ receives a \texttt{PositionTaken} as a response to its \texttt{ClaimPosition}. This case may happen if $q$ started travelling toward the destination when it was too far to perceive the previous \texttt{ClaimPosition} and \texttt{PositionTaken} messages.

In the first case, $q$ stops moving and sends an \texttt{InfoStopped} message, to advertise
its new position to the neighborhood, and starts a timeout. Snapped sensors receiving the
\texttt{InfoStopped} message, verify if the sender is inside their hexagons and in this case reply with a \texttt{IAYS} message (I Am Your Snapped), containing the sender and the receiver ID. If the stopped sensor receives a \texttt{IAYS} reply within the timeout, it sets its status to slave. Otherwise, if the timeout expires, it sets its status to free, not belonging to any hexagon.

In the second case, sensor $q$ compares its timestamp with the one included in the \texttt{ClaimPosition} message.
The sensor with lower timestamp wins the competition for the destination
and proceeds its travel, sending a \texttt{PositionTaken} message, while the other sensor waits for the arrival of the \texttt{IAS} message of the new snapped sensor to switch its status to slave.

In the third case, sensor $q$ sets its state to slave of the newly snapped sensor.
Notice that this timestamp based conflict is designed to avoid redundant replies to
\texttt{ClaimPosition} messages.

Figure \ref{fig:conflict} shows a typical conflict resolution scenario, where two sensors $r$ and $q$ receive a \texttt{SIP} message for the same position $x$ from two different snapped sensors.
Both $r$ and $q$ start travelling towards the destination $x$. Sensor $q$ reaches distance $d$ from the destination before sensor $r$, and sends a \texttt{ClaimPosition} message, with its timestamp. Sensor $r$ receives such message while travelling, and consequently stops
because the contention for position $x$ was won by sensor $q$.
Sensor $r$ sends an \texttt{InfoStopped} message to alert its neighborhood of its new position and starts a timeout.
In the case depicted in Figure \ref{fig:conflict}, $r$ stops inside the hexagon centered in position $x$.
For this reason, no snapped sensor replies to the \texttt{InfoStopped} message, thus after the timeout expiration, sensor $r$  switches its status to free.
After the expiration of the contention timeout, sensor $q$ broadcasts a \texttt{PositionTaken} message and switches to the snap status while definitely travelling to position $x$.
When $q$ reaches position $x$, it starts a neighbor discovery by sending a \texttt{IAS} message, in consequence of which, $r$ switches its status to slave.

\begin{figure}[t]
\begin{center}
\scalebox{0.44}{\includegraphics[]{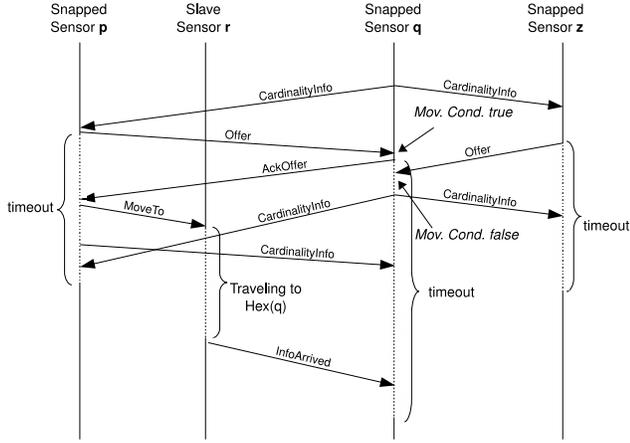}}
\end{center}
\caption{A typical scenario of the push activity} \label{fig:push}
\end{figure}

\section{P\&P: push activity}\label{sec:push}
To describe the push activity we distinguish the behavior of snapped and slave sensors and illustrate the role exchange mechanism introduced to uniform the energy consumption.

\subsection{Behavior of snapped sensors}

\subsubsection{Push proposal}
As soon as a snapped sensor $p$ terminates the snap activity,
it sends a \texttt{CardinalityInfo} message to its neighborhood.
Such message contains its ID and its  virtual cardinality.
Neighbor snapped sensors that receive this message update their information regarding sensor $p$ and evaluate
the opportunity to move slave sensors to its hexagon.

Even sensor $p$ evaluates the opportunity to move some of its slaves to adjacent hexagons to uniform the distribution of redundant sensors.
To this end, it uses its information regarding the neighbor snapped sensors, collected in the neighbor discovery phase.
Sensor $p$ looks for neighbor snapped sensors whose hexagons
verify the Moving Condition and have minimal cardinality.
Among these, it selects the closest, to which it sends an \texttt{Offer} message containing its
 virtual cardinality, and an identifier of the current transaction, (transaction ID).

If no sensor verifies the Moving Condition with $p$, sensor $p$ waits for further events.

\subsubsection{Push agreement}
The snapped sensor $q$ that receives an \texttt{Offer} message from $p$, verifies the validity of the Moving Condition as it
could have more updated information than $p$.
This way the responsibility of the slave movement is held by the receiver, thus ensuring that it only happens
when the Moving Condition is actually valid. This is particularly important to guarantee the algorithm termination.

Two cases may occur: 1) $q$ accepts the offer it received from $p$, or 2) $q$ leaves the offer unreplied.

In the first case, $q$ replies to $p$ with an \texttt{AckOffer} message, containing only the recipient and sender ID. Sensor $q$
updates its virtual cardinality value, advertising the new value to its snapped neighbors, with a \texttt{CardinalityInfo} message.
This way $q$ can participate in further operations of distribution of redundant sensors
with updated information and impede other snapped sensors to send  unnecessary offers.
When $q$ accepts an offer, it starts a timeout identified by the transaction ID received in the \texttt{Offer} message.
If $q$ does not receive any message within the timeout, containing the related transaction ID,
it decreases its virtual cardinality and advertises this change with a new \texttt{CardinalityInfo} message. This way the protocol is robust to possible node failures during the push activity.

The second case occurs if sensor $q$ verifies that the Moving Condition is unsatisfied with respect to sensor $p$.  For this reason it does not reply to the offer, causing the expiration of the offer timeout, after which sensor $p$ will be available to be engaged in other push actions.
This situation may happen when $p$ sends an offer on the basis of an outdated value of the virtual cardinality of $q$. As an example, the virtual cardinality can be outdated because in the meanwhile, $q$ has been involved in the push activity with other sensors.

\subsubsection{Selection of the sensor to push}
Sensor $p$ selects a slave $r$ to push and sends it a \texttt{MoveTo} message containing the sender and receiver ID, the position and the ID of the destination snapped node (in this case sensor $q$), and the transaction ID.
This selection is based on an energy saving criterion. Sensor $p$ selects the slave sensor $r$ that will remain with higher energy after the completion of the entire movement.

\subsection{Behavior of a slave sensor}
The slave sensor $r$ selected by sensor $p$ receives a \texttt{MoveTo} message and starts moving towards the hexagon of sensor $q$.  As soon as sensor $r$ crosses the boundary of the hexagon of $q$, it sends an \texttt{InfoArrived} message to $q$ and stops moving. The \texttt{InfoArrived} message contains the sender and receiver ID, the transaction ID, and the energy level of the sender.

\subsection{Role exchange}

The \HC\  algorithm provides that slaves and snapped sensors may occasionally
exchange their roles in order to balance the energy consumption over the set of available sensors.
Any time a slave $r$ has to make a movement across a hexagon as a consequence of a push action, it
sends a role exchange proposal consisting in a \texttt{Subst} message to the snapped sensor $p$ of the hexagon it is traversing, and starts a substitution timeout.

\texttt{Subst} messages contain the ID of sender and receiver, the energy level of the sender and the destination  coordinates.
The snapped sensor $p$ uses the energy level value of $r$ to decide if a role exchange
may be of benefit in balancing the overall energy consumption between the two sensors.
In this case, $p$ replies with an \texttt{AckSubst} message.

If sensor $r$ receives an \texttt{AckSubst} message within the substitution timeout,
it travels toward the snap position held by sensor $p$, while $p$ waits for the arrival of sensor $r$ before starting to travel
towards the destination initially targeted by $r$.
Sensor $r$ advertise its arrival to sensor $p$ with a \texttt{SubstArrival} message containing the same fields of the \texttt{AckSubst} message.
Sensor $p$ replies to $r$ with a \texttt{ProfilePacket} message that is necessary to enable a complete role exchange and starts travelling towards the destination.

If sensor $r$ does not receive an \texttt{AckSubst} message within the substitution timeout, it continues its travel towards the destination.

Slave and snapped sensor substitutions may also occur at the beginning of the slave travel.
In this case the substitution is started by the snapped sensor itself which already has all the available information
to evaluate the opportunity to perform the role exchange. Under these circumstances, the snapped sensor $p$ sends
a \texttt{MoveToSubst} message
containing the profile information necessary to perform the substitution.
As soon as sensor $r$ arrives in proximity to the snap position held by $p$, it sends the
\texttt{SubstArrival} message described before, after which $p$ starts travelling towards the destination.

\subsection{An example}
Figure \ref{fig:push} depicts a typical scenario of the push activity. The snapped sensor $q$ broadcasts its virtual cardinality with a \texttt{CardinalityInfo} message. The snapped sensors $p$ and $z$ receive this message and verify the Moving Condition with the updated information received from $q$. As both $p$ and $z$ satisfy the condition, they send an \texttt{Offer} message to $q$. Notice that the \texttt{Offer} message always contains an updated value of the virtual cardinality of the sender. Since each node can offer at most one sensor at a time the virtual cardinality does not change until the offer timeout expires, or the receiver replies with an \texttt{AckOffer} message.
Sensor $q$ receives the \texttt{Offer} message from $p$ before the one sent from sensor $z$. It verifies  the validity of the  Moving Condition with the updated virtual cardinality of $p$, received in the \texttt{Offer} message. As the Moving Condition is still satisfied, $q$ replies with an \texttt{AckOffer} message, incrementing its virtual cardinality and broadcasting a \texttt{CardinalityInfo} message.

When node $q$ receives the \texttt{Offer} message from $z$ it verifies  the Moving Condition again. Note that $z$ sent this message on the basis of an old value of the virtual cardinality of $q$. Thus $q$ finds that, as a consequence of the transaction just concluded with sensor $p$, the Moving Condition is unsatisfied with respect to sensor $z$, and consequently it does not reply to the offerer. Sensor $z$ waits until the expiration of the offer timeout, after which it is able to be engaged in other push actions.

Sensor $p$ receives an \texttt{AckOffer} message from $q$, thus it selects $r$ within its slaves, and send it a \texttt{MoveTo} message.  Sensor $r$ moves towards the hexagon of $q$, and sends an \texttt{InfoArrived} message as soon as it arrives. Sensor $p$ sends a \texttt{CardinalityInfo} message containing the decreased value of its virtual cardinality.

\section{P\&P: pull activity}

In the present section we distinguish three possible roles of sensors involved in the pull activity.

A first role is the one of the sensor detecting a coverage hole in a neighbor location.
This is the starter of the pull activity, which alters its \emph{ord}  value to enable push actions from nearby hexagons and sends
related trigger notification messages.

The second role is the one of the neighbor snapped sensors which receive the trigger notification messages while not having available slaves to send. These sensors act as forwarder of the trigger messages in order to reach hexagons with redundant slaves
that can be moved (pushed) to fill the coverage holes.

The third role is performed by the snapped sensors which receive a trigger message when having
available slaves to push. These sensors are informed of the changed \emph{ord} value of the neighbor snapped sensors,
and can contribute to fill the coverage holes by pushing the available slaves in the proper direction.

Notice that multiple trigger notification messages may reach the same sensor while performing any of the three listed
roles. Such messages are queued and processed with a priority inversely proportional to the distance from
the coverage hole.

\subsection{Behavior of sensors detecting coverage holes}

A snapped sensor $p$, located in proximity of some vacant positions (i.e. $\textrm{VP}(p)\neq \emptyset$), terminates the snap activity when no more sensors are available in $L(p)$. To give start to the pull activity, sensor $p$ verifies if there is the possibility to attract sensors from its snapped neighbors. To this purpose, sensor $p$ checks the validity of the Moving Condition with respect to all its snapped neighbors.

If $p$ can not continue the snap activity nor receive any sensor from its snapped neighbors, it starts the pull activity. To this purpose sensor $p$ sets its \emph{ord} value to zero, and advertises this change by broadcasting a \texttt{HoleInfo} message containing its ID, a hop counter $h$, its updated \emph{ord} value, the vacant position coordinates, and a timeout $t_\texttt{out}$ which depends on the value of $h$ (notice that this information is redundant but is introduced to increase the algorithm efficiency).
By modifying its $ord$ value, sensor $p$ alters the current situation with respect to the Moving Condition, enabling a new push activity from neighbor hexagons.

The hop counter $h$ represents the forwarding horizon of the \texttt{HoleInfo} message, that is the distance to be traversed by this message, expressed in number of hexagons.
Initially $h$ is set to zero, thus the snapped sensors receiving a \texttt{HoleInfo} message only update their
information about the sender \emph{ord} value and do not forward this message. If no new slave reaches $Hex(p)$ within the given timeout $t_\texttt{out}$, sensor $p$ increases $h$ and broadcasts a new \texttt{HoleInfo} message.

The timeout $t_\texttt{out}$ is calculated on the basis of the hop counter $h$ as the time necessary for a sensor located $(h+1)$  hops apart to reach $Hex(p)$, that is $t_\texttt{out} = (h+1) \cdot 2 R_\texttt{s}/v$, where $v$ is the sensor speed.

\begin{figure}[t]
\begin{center}
\scalebox{0.5}{\includegraphics[]{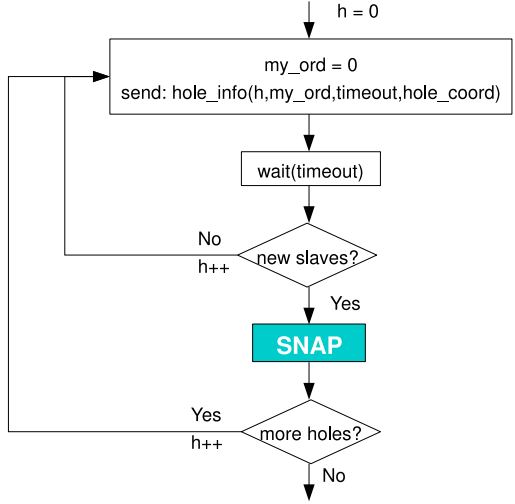}}
\end{center}
\caption{Behavior of a sensor detecting a coverage hole} \label{fig:pull_starter}
\end{figure}

Figure \ref{fig:pull_starter} illustrates the pull action performed by sensor $p$ as described above.

\subsection{Behavior of trigger forwarder sensors}
When a sensor $p$ receives a \texttt{HoleInfo} message and has not any slave to push toward the coverage hole,
it participates in the pull activity by forwarding this message when necessary.
In particular, it discards \texttt{HoleInfo} messages related to holes whose presence was already triggered by a snapped sensor $q$, unless they contribute additional information.
Indeed sensor $p$ evaluates new messages regarding a coverage hole previously advertised by sensor $q$ only if they come from:\\
1) snapped sensors with \emph{ord} value lower than $ord(p)$
or\\
2) snapped sensors with the same \emph{ord} as $p$ and hop counter $h$ which is higher than the forwarding horizon issued by sensor $q$.

Case 1) happens when a new snapped sensor $r$ detects the same coverage hole advertised by $q$, but the distance between $p$ and $r$ is lower than the distance between $p$ and $q$.
Case 2) happens when sensor $q$ issues a new hole trigger demanding a forwarding horizon extension.

When processing \texttt{HoleInfo} messages, sensor $p$ alters its status information,
and in particular sets its order value equal to the adjacent sender order value increased by 1. Sensor $p$ then forwards the trigger to its adjacent snapped sensors only if $h>0$. Such forwarded trigger message contains the updated status information of $p$ and a hop counter decreased by 1.

If sensor $p$ receives several \texttt{HoleInfo} messages concurrently, it inserts them in a pre-emptive priority queue, where
each message is treated with a priority that is inversely proportional to the distance from the coverage hole.

Sensor $p$ sets back its $ord$ to the original value as soon as the timeout of the \texttt{HoleInfo} message expires. This is necessary to stop the pull action after the coverage of the detected hole.

\subsection{Behavior of sensors pushing redundant slaves}

When a sensor $p$ receives a \texttt{HoleInfo} message and finds an available slave to push towards the coverage hole,
it updates the local information regarding its neighborhood.
Thanks to the sequence of order value alteration, $p$ finds
a valid Moving Condition with respect to the direction of the coverage hole and properly
starts a push activity.

\subsection{An example}
\begin{figure}[t]
\begin{center}
\scalebox{0.45}{\includegraphics[]{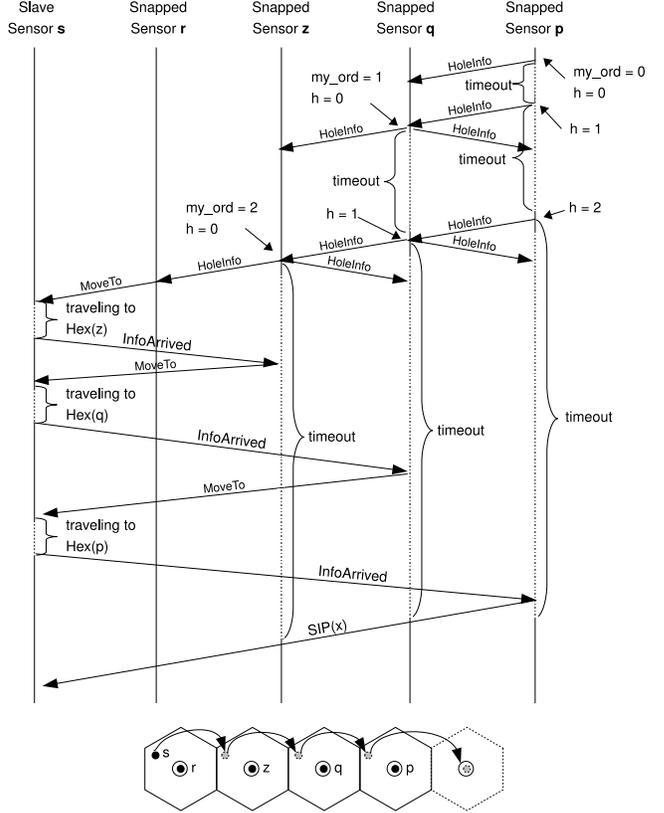}}
\end{center}
\caption{A typical scenario of the pull activity} \label{fig:pull_activity}
\end{figure}

Figure \ref{fig:pull_activity} shows a typical scenario of the pull activity. The snapped sensor $p$ detects a coverage hole in an adjacent position. Since $p$ has no slaves in its hexagon and the Moving Condition with respect to its neighbors is unsatisfied, it starts the pull activity by setting its $ord$ value to zero and broadcasting a \texttt{HoleInfo} message with null hop counter.
Since sensor $q$ does not have any slave to push toward $p$, at the expiration of the timeout, sensor $p$ broadcasts another  \texttt{HoleInfo} message increasing the previous hop counter. Sensor $q$ evaluates the hop counter of the \texttt{HoleInfo} message it received from $p$ and sets its own $ord$ value to 1. Sensor $q$ then forwards the trigger by broadcasting a \texttt{HoleInfo} message with decreased hop counter.
Once again the timeout set by $p$ expires because not even sensor $z$ has any slave to push, thus the procedure is repeated until the trigger, represented by the \texttt{HoleInfo} message, reaches sensor $r$ which instead has an available slave $s$ to push as it does according to the same procedure described in section \ref{sec:push}.

\section{P\&P: merge activity}


The fact that many sensors act as starters
implies the generation of  several tiling portions with different orientations.
The aim of the \HC\ algorithm is to cover the AoI with
a unique regular tiling thus minimizing overlaps of the sensing disks and
enabling a complete and uniform coverage.
Hence, the algorithm provides a merge mechanism to
be executed whenever a sensor $p$ receives a  neighbor
discovery message (\texttt{IAS}) from a snapped sensor $q$ belonging to another  tiling portion.

In this case, sensor $p$ chooses to join  the
oldest grid portion (it discriminates this situation by evaluating the
timestamp of the starter action, attached to any \texttt{IAS} message).

Notice that the detection of the sole neighbor
discovery messages is sufficient to ignite the
tiling merge activity because such messages are
sent after any tiling expansion and, if two tiling
portions come in radio proximity to each other, at least one of
them is increasing its extension.
In the following, we call $G_\texttt{old}$ and
$G_\texttt{new}$ the tiling portions with lower and
higher timestamp, respectively.
We distinguish three possible cases.

\noindent 1) Sensor $p$ belongs to $G_\texttt{new}$ and receives a \texttt{IAS} message from $q$ belonging to $G_\texttt{old}$.
If sensor $p$ is a slave, it switches its state
to free or to slave of the sensor $q$
depending on their mutual distance.
Sensor $p$ proactively communicates its new state to its
neighborhood by sending either an \texttt{InfoFree} or an \texttt{InfoSlave} message.
From now on $p$ honors only
messages from $G_\texttt{old}$ and ignores
those from $G_\texttt{new}$.

This proactive communication of the new state of $p$ is needed to
advertise the presence of $G_\texttt{new}$ when
there is no message activity within
$G_\texttt{new}$ that is perceivable by the sensors in
$G_\texttt{old}$.
This way, the snapped sensor which $p$ belonged to
can properly update its slave set.

If $p$ is instead a snapped sensor, it can not
immediately switch to its new state because
of its leading role inside $G_\texttt{new}$ (e.g.
it leads the slave sensors in $S(p)$ and performs
push and pull activities).
Hence $p$ temporarily assumes a hybrid role: it advertises itself
as free/slave to the nodes of $G_\texttt{old}$ with an \texttt{InfoFree}/\texttt{InfoSlave} message and, at
the same time, keeps on behaving as
snapped node in $G_\texttt{new}$ until it
receives a movement command (\texttt{SIP} or \texttt{MoveTo} message)
   coming from
$G_\texttt{old}$.

If $p$ received a \texttt{SIP} or a \texttt{MoveTo} command, $p$
moves to the new snap position electing one of
its slave in $G_\texttt{new}$ as a substitute with a \texttt{MoveToSubst} message.
The selected slave should reply with a \texttt{SubstArrival} upon arrival to the
snap position, within a given timeout. If this timeout expires before the reception of such \texttt{SubstArrival}
message, $p$ selects a new slave to snap.
The process goes on until no more slaves are available. In this case
$p$ ceases its snapped role inside $G_\texttt{new}$ advertising  its departure to its neighbors in
$G_\texttt{new}$, broadcasting a \texttt{Retirement} message.
Upon reception of a \texttt{Retirement} message the snapped neighbors that were located in positions adjacent
to the one that $p$ just freed, keep into account the new vacant position starting new snap activities.
If otherwise, $p$ receives a \texttt{SubstArrival} on time, it ceases its snapped role in $G_\texttt{new}$ and
honors the commands issued by the snapped node in $G_\texttt{old}$.
\\
2) Sensor $p$ belongs to $G_\texttt{old}$ and receives a \texttt{IAS} message from $q$ belonging to $G_\texttt{new}$:
if $p$ is a slave it ignores all messages from $G_\texttt{new}$.
If $p$ is snapped, it performs a neighbor discovery sending a \texttt{IAS} message,
ignores all messages coming from $G_\texttt{new}$,
apart from the neighbor discovery replies, and honors only messages
from $G_\texttt{old}$.
Observe that the neighbor discovery is necessary to ignite the merge mechanism and allows each snapped sensor
in  $G_\texttt{old}$ to collect complete
information on nearby sensors that previously
belonged to  $G_\texttt{new}$.\\
3) Sensor $p$ is free: sensor $p$ honors only messages from
$G_\texttt{old}$ and ignores those from
$G_\texttt{new}$.

\section{Simulation results}
\label{sec:exp_setup}

In order to evaluate the performance of the P\&P protocol
we developed a simulator on the basis of the
wireless module of the  OPNET modeler software
\cite{opnet}.

The experimental activity required the definition of some setup parameters:
$R_\texttt{tx}=11$ m, $R_\texttt{s}=5$ m and the sensor speed is 1 m/sec.

\begin{figure}
 \centering
  \subfigure[]{\scalebox{0.25}{
\includegraphics[]{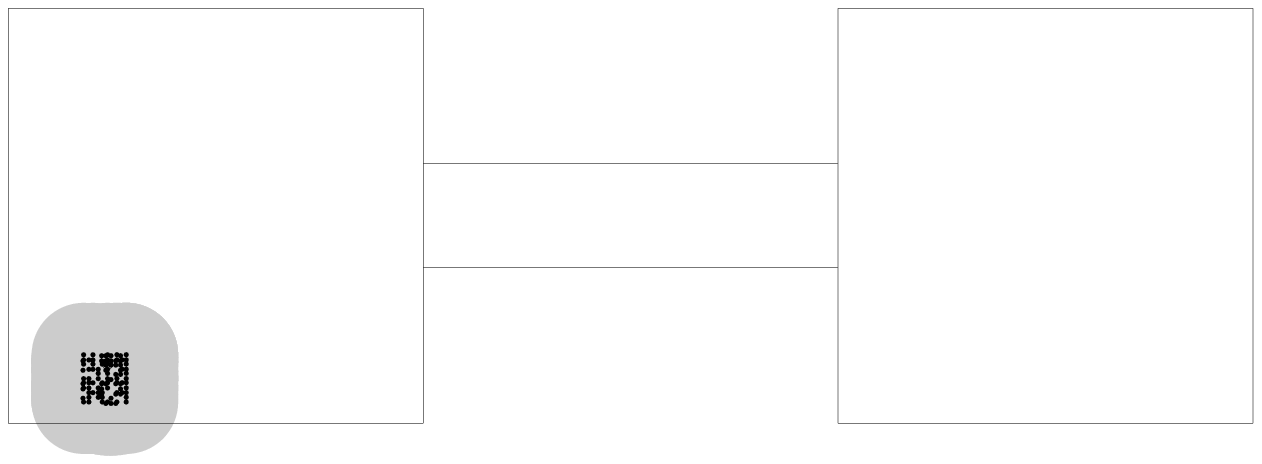}}}
\subfigure[]{\scalebox{0.25}{
\includegraphics[]{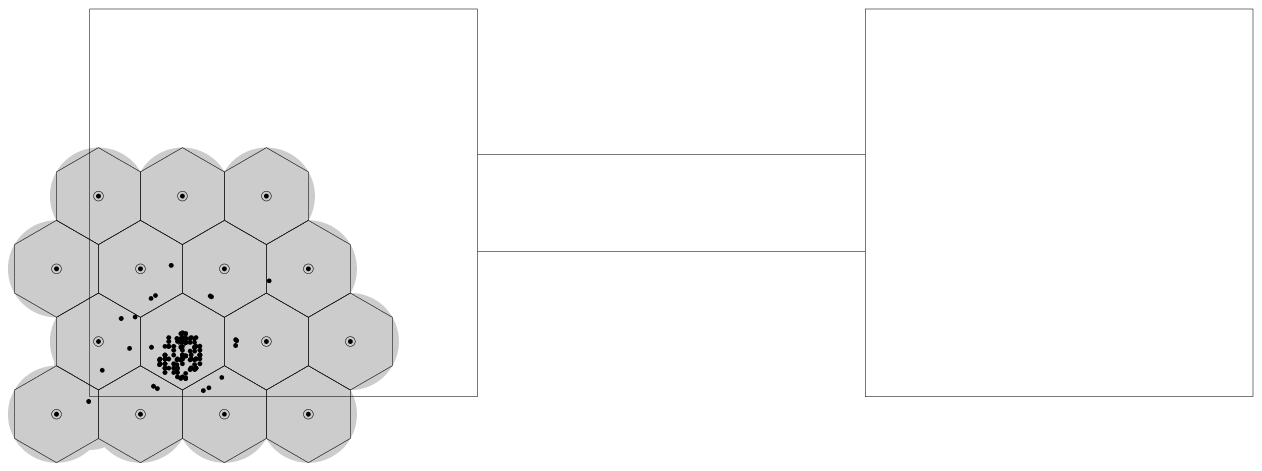}}}
\subfigure[]{\scalebox{0.25}{
\includegraphics[]{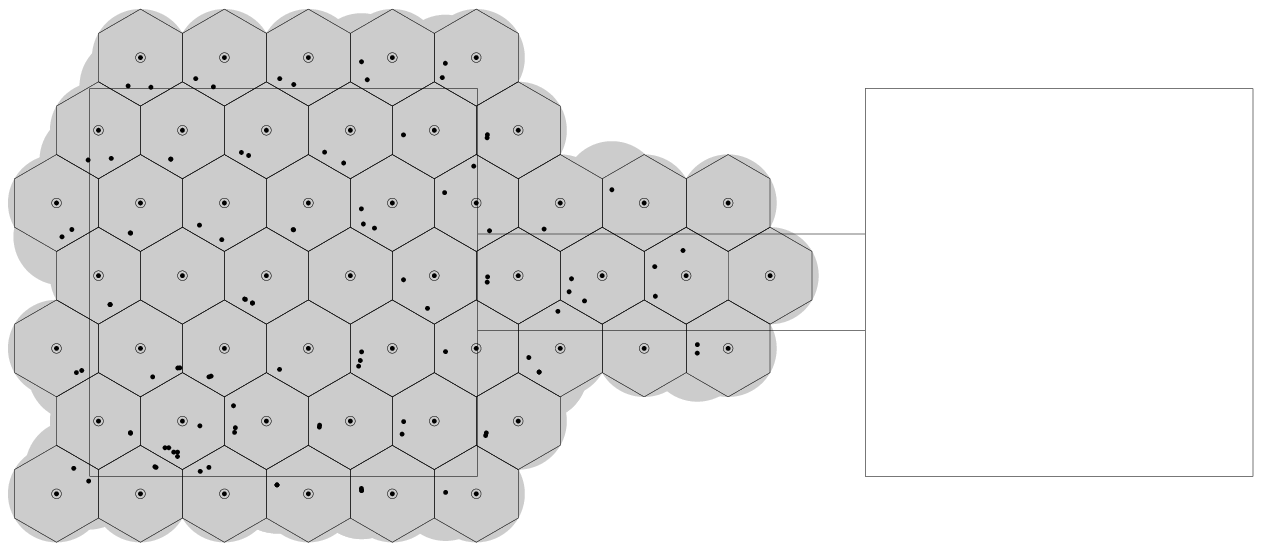}}}
\subfigure[]{\scalebox{0.25}{
\includegraphics[]{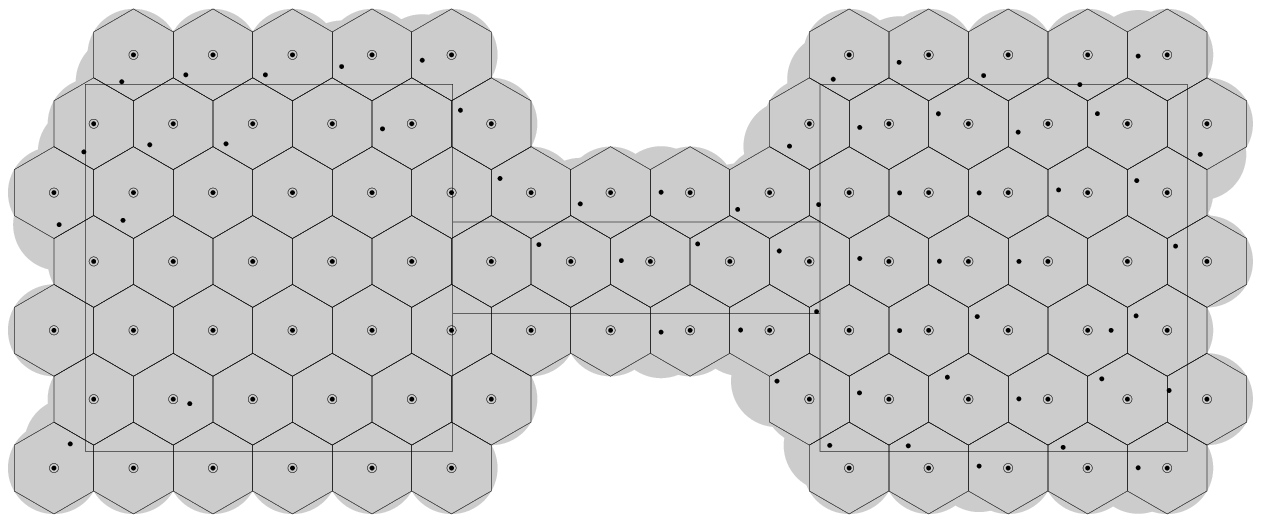}}}
\caption{Coverage of an irregular AoI}
\label{fig:osso}
\end{figure}

We show some examples of final deployments provided by the proposed protocol.
Figure \ref{fig:osso} gives a synthetic representation of
how the sensor deployment evolves under P\&P
when starting with an initial configuration where
150
sensors are sent from a high density region.
The AoI has a complex shape in which a narrows connects two
square regions
40 m $\times$ 40 m.

\begin{figure}
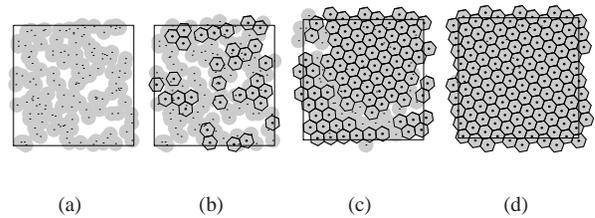

\begin{center}
  \subfigure[]{\scalebox{0.02}{
\includegraphics[]{./immagini/confronti/random/a.1}}}
\subfigure[]{\scalebox{0.02}{
\includegraphics[]{./immagini/confronti/random/b.1}}}
\subfigure[]{\scalebox{0.02}{
\includegraphics[]{./immagini/confronti/random/c.1}}}
\subfigure[]{\scalebox{0.02}{
\includegraphics[]{./immagini/confronti/random/d.1}}}
\end{center}
\caption{Deployment with random initial distribution} \label{fig:deployment_comparison_SS}
\end{figure}

In the following we also show three other deployment examples obtained with different initial sensor deployments over the
same AoI, that is a square 80 m $\times$ 80 m.

In the first example, the initial deployment evidences a random distribution of sensors over the AoI as depicted in
Figure \ref{fig:deployment_comparison_SS}(a). With such initial configuration the deployment obtained by P\&P evolves through the intermediate stages (b) and  (c), achieving the final deployment shown in figure (d).

\begin{figure}
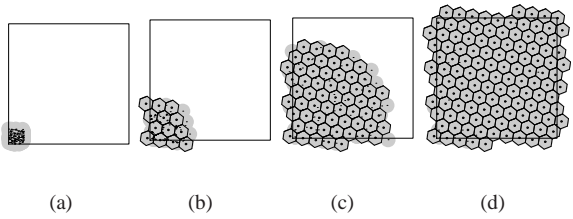

\begin{center}
  \subfigure[]{\scalebox{0.02}{
\includegraphics[]{./immagini/confronti/grumo_laterale/a.1}}}
\subfigure[]{\scalebox{0.02}{
\includegraphics[]{./immagini/confronti/grumo_laterale/b.1}}}
\subfigure[]{\scalebox{0.02}{
\includegraphics[]{./immagini/confronti/grumo_laterale/c.1}}}
\subfigure[]{\scalebox{0.02}{
\includegraphics[]{./immagini/confronti/grumo_laterale/d.1}}}
\end{center}
\caption{Deployment with high density initial distribution at the boundaries of the AoI} \label{fig:deployment_comparison_GL}
\end{figure}

In the second example
the initial deployment consists of a high density region at the boundaries of the AoI as depicted in
Figure \ref{fig:deployment_comparison_GL}(a). With such initial configuration P\&P achieves the final deployment detailed in (d) evolving as in (b) and (c).

\begin{figure}
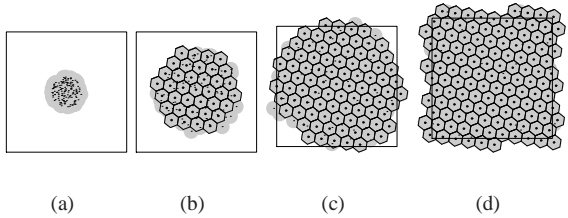

\begin{center}
  \subfigure[]{\scalebox{0.02}{
\includegraphics[]{./immagini/confronti/grumo_centrale/a.1}}}
\subfigure[]{\scalebox{0.02}{
\includegraphics[]{./immagini/confronti/grumo_centrale/b.1}}}
\subfigure[]{\scalebox{0.02}{
\includegraphics[]{./immagini/confronti/grumo_centrale/c.1}}}
\subfigure[]{\scalebox{0.02}{
\includegraphics[]{./immagini/confronti/grumo_centrale/d.1}}}
\end{center}
\caption{Deployment with high density initial distribution at the center of the AoI} \label{fig:deployment_comparison_GC}
\end{figure}

In the third example,
the initial deployment consists of high density region at the center of the AoI as depicted in
Figure \ref{fig:deployment_comparison_GC}(a). With such initial configuration, the deployment obtained by P\&P evolves through the configurations shown in (b) and (c), reaching the final deployment given in (d).
\begin{figure}[h]
\begin{center}
\begin{minipage}{0.5\columnwidth}
\centering
\includegraphics[width = 0.7\columnwidth, angle=-90]{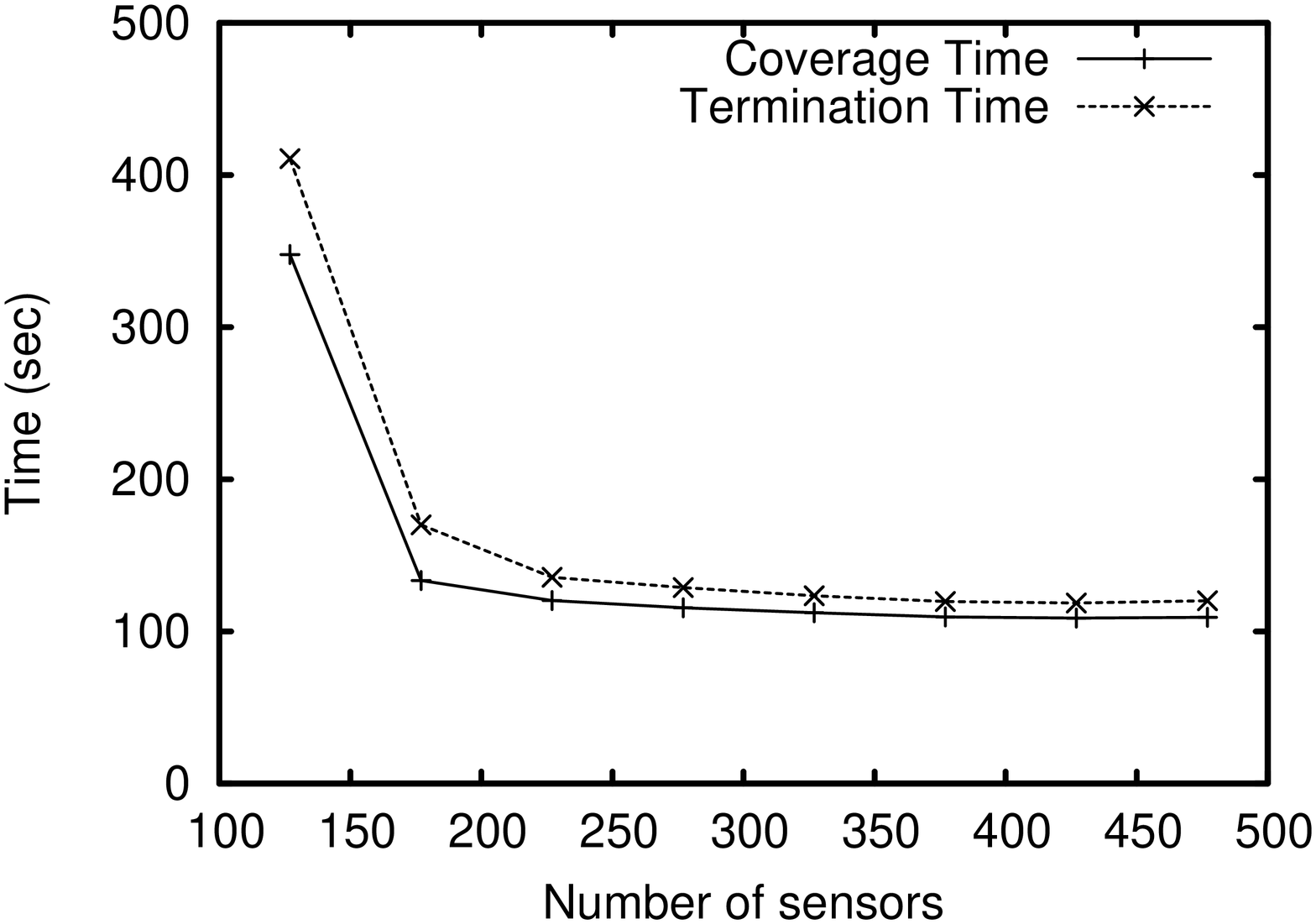}
\caption{Term. and coverage time} \label{fig:pp.times}
\end{minipage}
\begin{minipage}{0.5\columnwidth}
\centering
\includegraphics[width = 0.7\columnwidth, angle=-90]{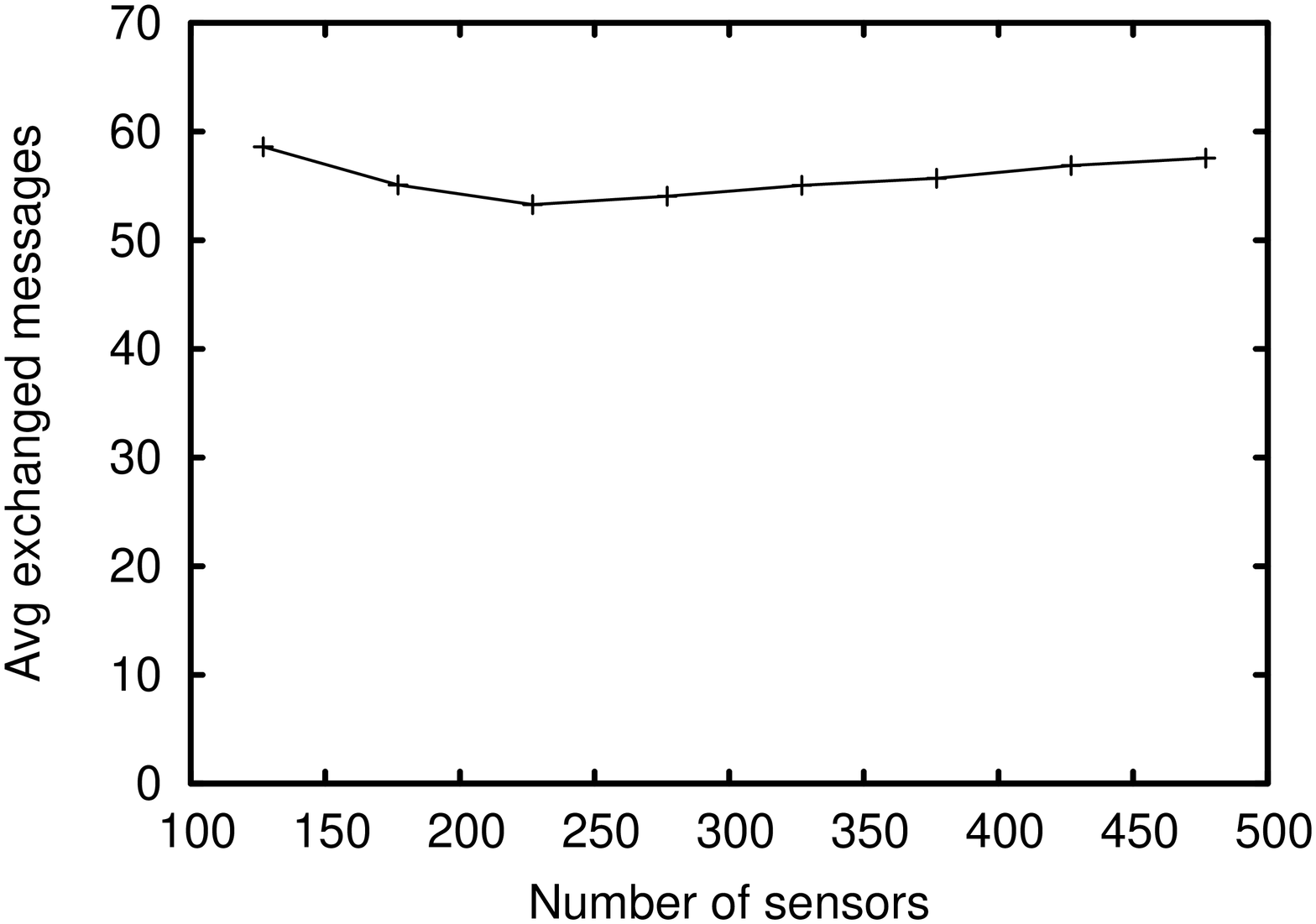}
\caption{Nr of message exchanges} \label{fig:pp.nr_mess}
\end{minipage}
\\
\begin{minipage}{0.5\columnwidth}
\centering
\includegraphics[width = 0.7\columnwidth, angle=-90]{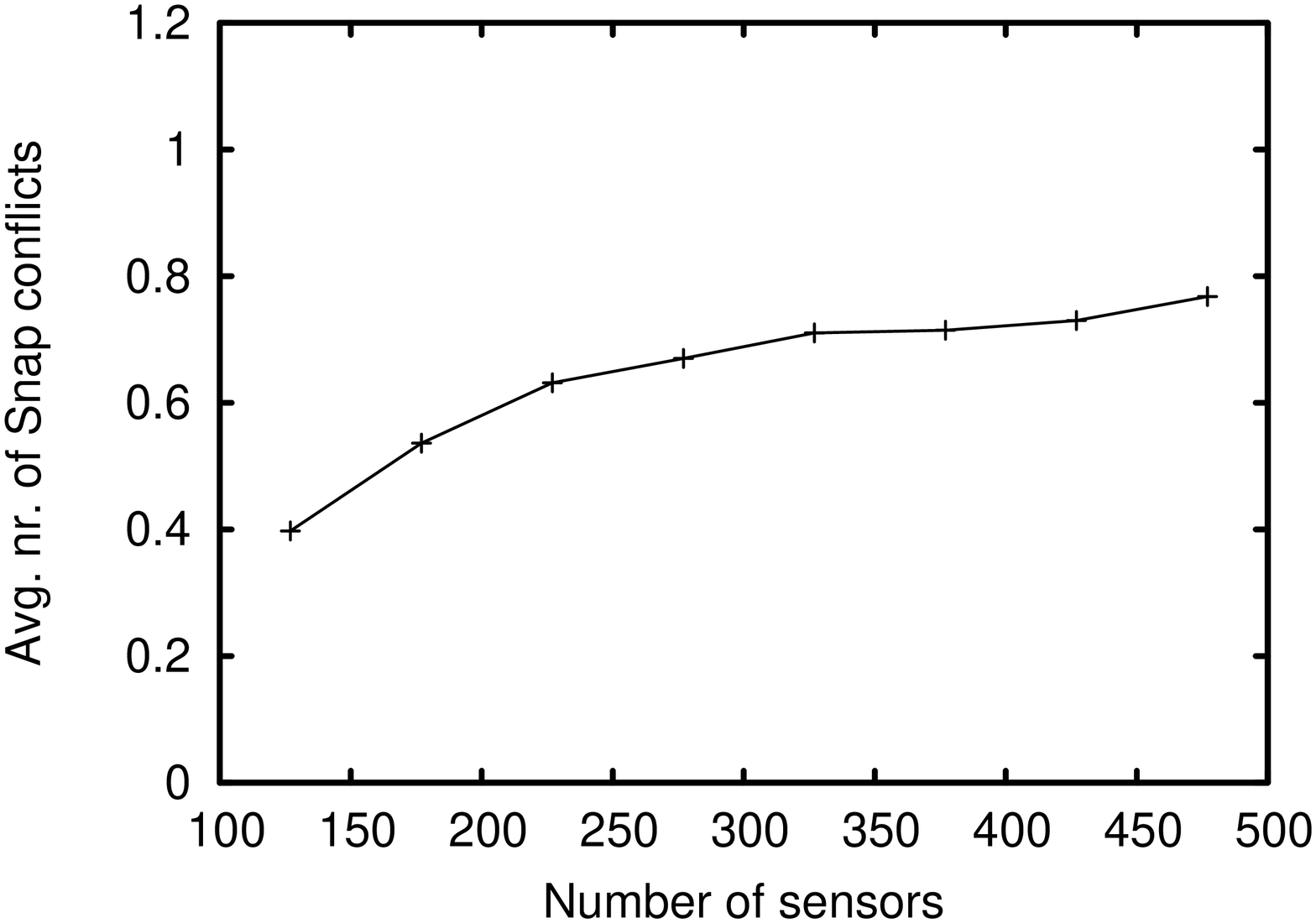}
\caption{Snap conflicts} \label{fig:pp.snap_conflicts}
\end{minipage}
\begin{minipage}{0.5\columnwidth}
\centering
\includegraphics[width = 0.7\columnwidth, angle=-90]{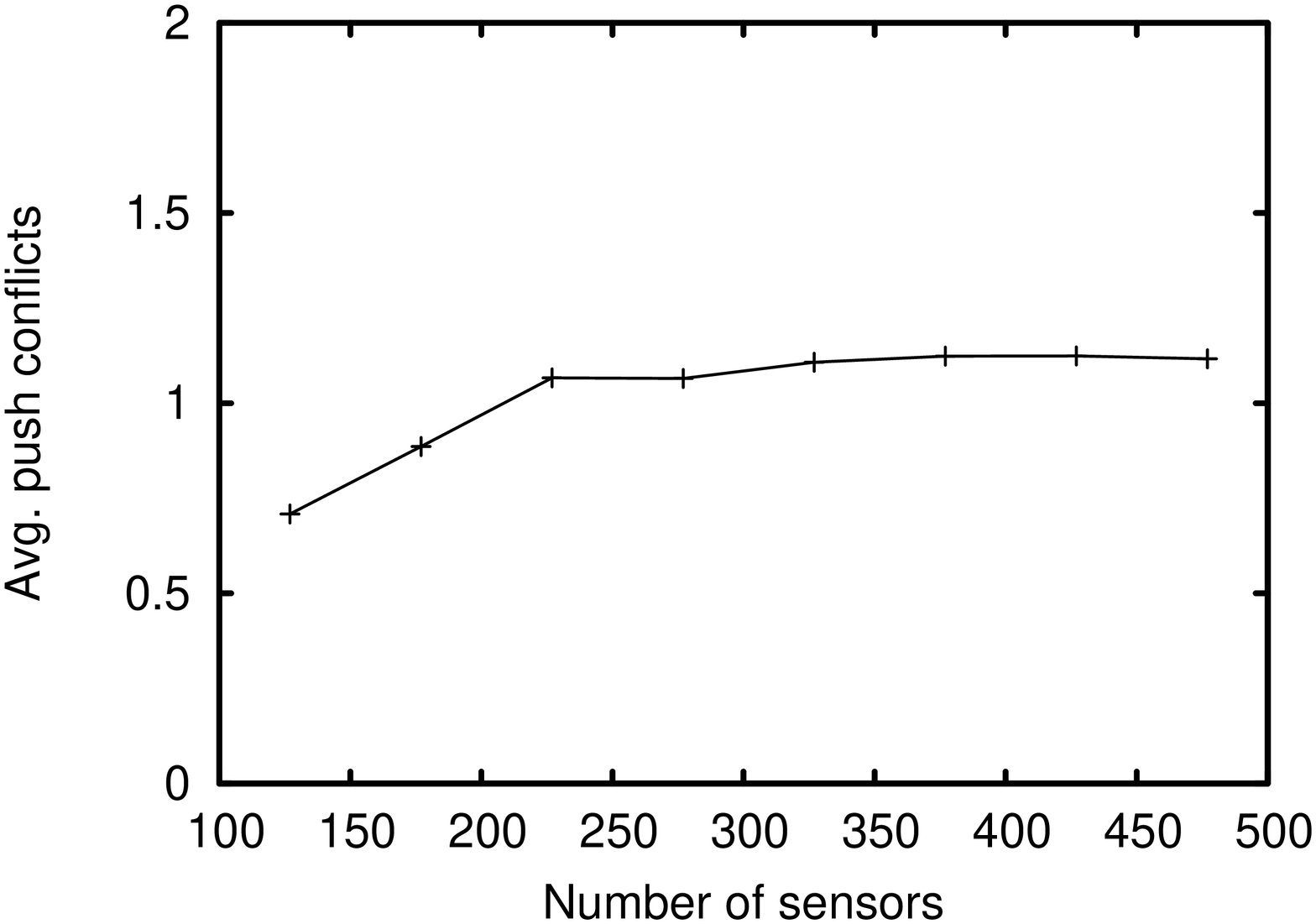}
\caption{Push conflicts} \label{fig:pp.push_conflicts}
\end{minipage}
\end{center}
\end{figure}

We introduce some experiments conducted starting from the configuration depicted in figure \ref{fig:deployment_comparison_GC}(a), by varying the number of sensors. Results are average values calculated over 30 simulation runs.
Figure \ref{fig:pp.times} shows the coverage and termination time of the protocol execution.
By coordinating distributed decisions and solving local conflicts, the P\&P protocol guarantees the termination of the \HC\ algorithm in moderate time. Notice that after the coverage completion, the \HC\ algorithm keeps on regulating some movements to uniform the redundant sensor density. The termination time evidences the capability of the P\&P protocol to reach a final stable configuration, where neither movements nor message exchanges are performed.

The average number of message exchanges, shown in Figure \ref{fig:pp.nr_mess}, evidences a good scalability of the P\&P protocol. Indeed this number remains stable even when the number of sensors increases significantly.

Figure \ref{fig:pp.snap_conflicts} represents the number of conflicting snap actions, described in section \ref{sec:snap}, averaged over the number of snap positions.
The asynchronous behavior of P\&P guarantees the resolution of the few position conflicts that arise as a consequence of the distributed execution of the algorithm \HC. Although growing with the number of available sensors, the average number of snap conflicts remains significantly smaller than 1, meaning that, in the considered scenarios, no more than one conflict happens per snap position.

A push conflict happens when a push offer made by one sensor becomes obsolete in consequence to push actions performed by others. Despite the distributed execution of the P\&P protocol, the average number of push conflicts per slave sensor is stable with a growing number of sensors, as shown in Figure \ref{fig:pp.push_conflicts}.

\section{Conclusions}

In this paper we introduce P\&P, a communication protocol that permits a correct and efficient coordination of sensor movements in agreement with the \HC\ algorithm.

Unlike previous works which introduce deployment algorithms without
formalizing the related protocol, we address the realistic applicability of this approach.
Indeed we deeply investigate and propose protocol solutions to the possible conflicts that
 may arise when   asynchronous local decisions are to be coordinated.

Simulation results show the performance of our protocol under a
range of operative settings, including conflict situations,
irregularly shaped target areas, and node failures. These results
evidence the protocol capabilities to fulfill the algorithm
requirements and in particular termination, completeness and
stability of the final coverage.

\bibliographystyle{IEEEtran}
\bibliography{IEEEabrv,bibliografia}
\end{document}